# Experimental evidence for non-Abelian gauge potentials in twisted graphene bilayers

Long-Jing Yin[1,2], Jia-Bin Qiao[1,2], Wei-Jie Zuo[1,2], Wen-Tian Li[1,2], and Lin He[1,2,*]

[1] Department of Physics, Beijing Normal University, Beijing, 100875, People's Republic of China
[2] The Center of Quantum Studies, Beijing Normal University, Beijing, 100875, People's Republic of China

* Email: helin@bnu.edu.cn

**The methods for realizing of non-Abelian gauge potentials have been proposed in many different systems in condensed matter[1-5]. The simplest realization among them may be in a graphene bilayer obtained by slightly relative rotation between the two layers[4]. Here we report the experimental evidence for non-Abelian gauge potentials in twisted graphene bilayers by scanning tunnelling microscopy and spectroscopy. At a magic twisted angle, $\theta \approx (1.11 \pm 0.05)°$, a pronounced sharp peak, which arises from the nondispersive flat bands at the charge neutrality point, are observed in the tunnelling density of states due to the action of the non-Abelian gauge fields[4,6-8]. Moreover, we observe confined electronic states in the twisted bilayer, as manifested by regularly spaced tunnelling peaks with energy spacing $\delta E \approx v_F/D \approx 70$ meV (here $v_F$ is the Fermi velocity of graphene and $D$ is the period of the Moiré patterns). Our results direct demonstrate that the non-Abelian gauge potentials in twisted graphene bilayers confine low-energy electrons into a triangular array of quantum dots following the modulation of the Moiré patterns.**

Because of Klein tunnelling[9,10], it is difficult to confine Dirac fermions in graphene. Therefore, the confinement of electrons in graphene has attracted much attention over the years, not least because the requirement of potential applications in graphene-based devices[11-14]. Recently, a novel recipe for confining the Dirac fermions by using twisted graphene bilayers has been suggested[4,6-8]. It was predicted that electrons in the twisted bilayers can be tuned from chiral Dirac fermions to localized electrons by simply varying the rotation angle. The modulations in the interlayer hopping of twisted bilayers generate an effective non-Abelian gauge potential $\hat{\mathbf{A}}$ in the low-energy electronic theory[4] and result in nondispersive flat bands at charge neutrality point for a discrete set of magic angles[6-8]. Experimentally, electronic properties of twisted graphene bilayers have been studied by several groups[15-22]. It was demonstrated explicitly that the energy spacing of two low-energy van Hove singularities (VHSs) $\Delta E_{VHS}$, which originate from the two saddle points in the band structures of the twisted graphene bilayer[23], decreases with the twisted angle[15,18,20-22] and reduces to only $\Delta E_{VHS} \approx 12$ meV for the sample with $\theta \approx (1.16 \pm 0.04)°$ [15]. However, a systematic experimental verification of the nondispersive flat bands in the twisted bilayers is still lacking, owing to the difficulty to obtain the twisted bilayers with small rotation angles, whereas, the discrete magic angles are predicted to be smaller than 1.5°.

Our scanning tunneling microscopy (STM) and spectroscopy (STS) measurements were carried out on highly oriented pyrolytic graphite (HOPG) surface (see Supplementary Information for method of STM measurements). The surface few-layer of HOPG usually decouples from the bulk[15,21,24-27] and, more importantly, the twisted graphene bilayers with various rotation angles, especially the small rotation angles, can be easily observed on HOPG surface[21]. Therefore, the surface of HOPG provides a natural ideal platform to verify the fundamental prediction: the existence of non-Abelian gauge potentials in twisted graphene bilayers[4].

Figure 1 shows representative STM images of several twisted graphene bilayers with different twisted angles, which can be derived from the period of the Moiré patterns $D$, using $D = a/[2\sin(\theta/2)]$ with the graphene lattice constant $a \sim 0.246$ nm[15-22]. Because of the twist, the local stacking orders of the two layers depend on positions in the Moiré patterns, as shown in Fig. 1e. Our experimental results demonstrate that the bright dots in the STM images are AA-stacking regions (all atoms of the upper layer sitting on top of the atoms of the lower layer), which surround with AB-stacking regions (a C atom of one layer sits in the center of a graphene hexagon of the other layer). The corrugation of the Moiré patterns measured in the STM images depends sensitively on the bias voltages between the STM tip and the sample (see Supplementary Fig. S1 for details of measurements), as shown in Fig. 1f. It indicates that spatial variation of local density of states (DOS) is the main source of the observed periodic corrugation in the twisted bilayers.

The local DOS of the twisted bilayers was measured by STS at 4.3 K. Figure 2a-2d shows representative differential conductance spectra measured on four twisted bilayers with different twisted angles. The energy spacing of the two VHSs is measured as $\Delta E_{VHS} \approx 350$ meV for the 3.1° sample (Fig. 2d) and it reduces to only about 80 meV for the 1.88° bilayer (Fig. 2c). The decrease of the $\Delta E_{VHS}$ from the 3.1° bilayer to the 1.88° bilayer agrees quite well with that reported previously[15,18,20-22] and is also consistent with the theoretical result shown in Fig. 3a. For the 1.11° and 0.88° bilayers, the two VHSs merge into a pronounced sharp peak located at the charge neutrality point in the tunnelling spectra, as shown in Fig. 2a and Fig. 2b. This quite differs from the bilayers with larger twisted angles. Even for the sample with $\theta \approx$ 1.16°, there are two peaks with energy spacing of 12 meV in its low-energy DOS[15]. Additionally, the FWHM (full-width at half-maximum) of the sharp peak in the 1.11° bilayer is measured to be only 18 meV, which is much smaller than that, ~ 40 meV, of the 1.16° bilayer[15]. Therefore, the sharp DOS peak at the charge neutrality point, as shown in Fig. 2b, is clearly a sign of the presence of low-energy flat bands in the 1.11° bilayer (see Fig. 3a for theoretical result). Theoretically, the first magic angle for the appearance of the non-dispersion flat bands generated by the non-Abelian gauge potentials $\hat{\mathbf{A}}$ is predicted to be around 1.0°-1.5° in the twisted bilayers (the exact value of the first magic angle depends upon the parametrization models)[4,6-8]. Our experimental result indicates that the first magic angle should be at about 1.11°.

With further lowering the twisted angle from 1.11° to 0.88°, the FWHM of the DOS peak increases from 18 meV to about 30 meV, as shown in Fig. 2a. Such a behavior is reasonable with considering the fact that the lowest sub-band in twisted bilayer becomes dispersive once more, i.e., the bandwidth increases, before collapsing to flat bands again at the second magic angle[8], ~ 0.5°. The flat bands generated by the non-Abelian gauge potentials are predicted to be confined mainly in the AA-stacking regions (the bright dots in the STM images)[4]. This has been demonstrated explicitly by the spatial resolved STS spectra, as shown in Fig. 2b. The d$I$/d$V$ maps shown in Fig. 2e-2g further demonstrated directly the strong localization of the flat bands in the bright dots of the twisted bilayer. Our d$I$/d$V$ maps, as shown in Fig. 2h-2j, also indicate that there is a strong density modulation, characteristic of charge-density waves[15], for electrons with energies close to the VHSs in twisted bilayers with $\theta < 2°$.

Besides the pronounced sharp peak at the charge neutrality point, a series of discrete tunnelling peaks appear in the STS spectra recorded in the 1.11° bilayer, as shown in Fig. 2b and Fig. 3b. The energy spacing δ$E$ of these peaks is almost regularly spaced with the averaged value as about 70 meV (Fig. 3b). These DOS peaks may arise from Moiré Bloch bands generated by the Moiré pattern periodicity[8]. The expected energy separation for the confined electronic states in the 1.11° Moiré pattern can be estimated as δ$E \approx v_F/D \approx 70$ meV ($v_F$ is the Fermi velocity of graphene), which is consistent well with our experimental result. This result, together with the d$I$/d$V$ maps shown in Fig. 2, demonstrated directly that the non-Abelian gauge potentials confine low-energy electrons in twisted graphene bilayers into a triangular array of quantum dots following the modulation of the Moiré patterns.

The ability to tune electrons from Dirac fermions to localized electrons, or vice versa, by simply varying the rotation angle in the twisted bilayers can be further demonstrated by measuring the tunnelling spectra in high magnetic fields. Figure 4 shows the high magnetic fields tunnelling spectra of two representative twisted bilayers. For small twisted angles, such as the 1.11° and 0.88°, the electrons are strongly localized and the picture of two-dimensional free electron gas no longer applies in the twisted bilayers. Therefore, we cannot observe Landau-level quantization in these samples, as shown in Fig. 4a (see Supplementary Fig. S2 and Fig. S3 for more experimental results). The increase of the FWHM of the peak at the charge neutrality point with increasing the magnetic fields, as shown in Fig. 4c, may arise from a competition between twist-induced localization and cyclotron motion generated by the magnetic fields[16]. The magnetic length generated by the magnetic field of 7 T is estimated to be $l_B = \sqrt{\hbar/eB}$ ≈ 10 nm, which is comparable to the period of the Moiré patterns. For the 3.1° bilayer, the low-energy band structure (within the two VHSs) can be described well by two Dirac cones separated in reciprocal space[23], as shown in Fig. 3a. Consequently, it is expected to observe Landau quantization of massless Dirac fermions in twisted bilayers with larger rotation angles[16,28]. This is demonstrated explicitly in our experiment, where the observed Landau levels sequence in the 3.1° bilayer, as shown in Fig. 4b and Fig. 4d, depends on the square-root of both level index *n* and magnetic field *B*.

In summary, we demonstrated that the combination of two graphene sheets with a stacking fault leads to novel and exotic properties not present in graphene monolayer. Because of the effective non-Abelian gauge fields, the rotation angle could transfer the charge carriers in the twisted bilayers from massless Dirac fermions into well localized electrons, or vice versa, efficiently. This provides a new route to tune the electronic properties of graphene systems, which will be essential in future graphene nanoelectronics.

**REFERENCES:**

1. Lin, Y.-J., Jimenez-Garcia, K., Spielman, I. B. Spin-orbit-coupled Bose-Einstein condensates. *Nature* **471**, 83-88 (2011).

2. Gonzalez, J., Guinea, F., Vozmediano, M. A. H. Continuum approximation to fullerene molecules. *Phys. Rev. Lett.* **69**, 172 (1992).

3. Osterloh, K., Baig, M., Santos, L., Zoller, P., Lewenstein, M. Cold atoms in non-Abelian gauge potentials: from the Hofstadter "moth" to lattice gauge theory. *Phys. Rev. Lett.* **95**, 010403 (2005).

4. San-Jose, P., Gonzalez, J., Guinea, F. Non-Abelian gauge potentials in graphene bilayers. *Phys. Rev. Lett.* **108**, 216802 (2012).

5. Gopalakrishnan, S., Ghaemi, P., Ryu, S. Non-Abelian SU(2) gauge fields through density wave order and strain in graphene. *Phys. Rev. B* **86**, 081403(R) (2012).

6. Suarez Morell, E., Correa, J. D., Vargas, P., Pacheco, M., Barticevic, Z. Flat bands in slightly twisted bilayer graphene: tight-binding calculations. *Phys. Rev. B* **82**, 121407(R) (2010).

7. Trambly de Laissardiere, G., Mayou, D., Magaud, L. Localization of Dirac electrons in rotated graphene bilayers. *Nano Lett.* **10**, 804-808 (2010).

8. Bistritzer, R., MacDonald, A. H. Moire bands in twisted double-layer graphene. *PNAS(USA)* **108**, 12233-12237 (2011).

9. Katsnelson, M. I., Novoselov, K. S., Geim, A. K. Chiral tunneling and the Klein paradox in graphene. *Nature Phys.* **2**, 620-625 (2006).

10. He, W.-Y., Chu, Z.-D., He, L. Chiral tunneling in a twisted graphene bilayer. *Phys. Rev. Lett.* **111**, 066803 (2013).

11. Ponomarenko, L. A., Schedin, F., Katsnelson, M. I., Yang, R., Hill, E. W., Novoselov, K. S., Geim, A. K. Chaotic Dirac billiard in graphene quantum dots. *Science* **320**, 356-358 (2008).

12. Calleja, F., Ochoa, H., Garnica, M., Barja, S., Navarro, J. J., Black, A., Otrokov, M. M., Chulkov, E. V., Arnau, A., Vazquez de Parga, A. L., Guinea, F., Miranda, R. Spatial variation of a giant spin-orbit effect induces electron confinement in graphene on Pb islands. *Nature Phys.* **11**, 43-47 (2015).

13. Phark, S.-h, Borme, J., Vanegas, A. L., Corbetta, M., Sander, D., Kirschner, J. Direct observation of electron confinement in epitaxial graphene nanoislands. *Nano Lett.* **5**, 8162 (2011).

14. Wang, X., Ouyang, Y., Jiao, L., Wang, H., Xie, L., Wu, J., Guo, J., Dai, H. Graphene nanoribbons with smooth edges behave as quantum wires. *Nature Nano.* **6**, 563 (2011).

15. Li, G., Luican, A., Lopes dos Santos, J. M. B., Castro Neto, A. H., Reina, A., Kong, J., Andrei, E. Y. Observation of Van Hove singularities in twisted graphene layers. *Nat. Phys.* **6**, 109-113 (2009).

16. Luican, A., Li, G., Reina, A., Kong, J., Nair, R. R., Novoselov, K. S., Geim, A. K., Andrei, E. Y.. Single-Layer Behavior and Its Breakdown in Twisted Graphene

Layers. *Phys. Rev. Lett.* **106**, 126802 (2011).

17. Yan, W., He, W.-Y., Chu, Z.-D., Liu, M., Meng, L., Dou, R.-F., Zhang, Y., Liu, Z., Nie, J.-C., He, L. Strain and curvature induced evolution of electronic band structures in twisted graphene bilayer. *Nature Commun.* **4**, 2159 (2013).

18. Yan, W., Liu, M., Dou, R.-F., Meng, L., Feng, L., Chu, Z.-D., Zhang, Y., Liu, Z., Nie, J.-C., He, L. Angle-dependent van Hove singularities in a slightly twisted graphene bilayer. *Phys. Rev. Lett.* **109**, 126801 (2012).

19. Ohta, T., Robinson, J. T., Feibelman, P. J., Bostwick, A., Rotenberg, E., Beechem, T. E. Evidence for Interlayer Coupling and Moiré Periodic Potentials in Twisted Bilayer Graphene. *Phys. Rev. Lett.* **109**, 186807 (2012).

20. Brihuega, I., Mallet, P., Gonzalez-Herrero, H., Trambly de Laissardiere, G., Ugeda, M. M., Magaud, L., Gomez-Rodriguez, J. M., Yndurain, F., Veuillen, J.-Y. Unraveling the Intrinsic and Robust Nature of van Hove Singularities in Twisted Bilayer Graphene by Scanning Tunneling Microscopy and Theoretical Analysis. *Phys. Rev. Lett.* **109**, 196802 (2012).

21. Yin, L.-J., Qiao, J.-B., Wang, W.-X., Chu, Z.-D., Zhang, K. F., Dou, R. F., Gao, C. L., Jia, J.-F., Nie, J.-C., He, L. Tuning structures and electronic spectra of graphene layers with tilt grain boundaries. *Phys. Rev. B* **89**, 205410 (2014).

22. Yan, W., Meng, L., Liu, M., Qiao, J.-B., Chu, Z.-D., Dou, R.-F., Liu, Z., Nie, J.-C., Naugle, D. G., He, L. Angle-dependent van Hove singularities and their breakdown in twisted graphene bilayers. *Phys. Rev. B* **90**, 115402 (2014).

23. Lopes dos Santos, J. M. B., Peres, N. M. R., Castro Neto, A. H. Graphene bilayer

with a twist: electronic structure. *Phys. Rev. Lett.* **99**, 256802 (2007).

24. Matsui, T. *et al.* STS Observations of Landau Levels at Graphite Surfaces. *Phys. Rev. Lett.* **94**, 226403 (2005).

25. Li, G., Andrei, E. Y. Observation of Landau levels of Dirac fermions in graphite. *Nat. Phys.* **3**, 623-627 (2007).

26. Xu, R., Yin, L.-J., Qiao, J.-B., Bai, K.-K., Nie, J.-C., He, L. Direct probing of the stacking order and electronic spectrum of rhombohedral trilayer graphene with scanning tunneling microscopy. *Phys. Rev. B* **91**, 035410 (2015).

27. Yin, L.-J., Li, S.-Y., Qiao, J.-B., Nie, J.-C., He, L. Landau quantization in graphene monolayer, Bernal bilayer, and Bernal trilayer on graphite surface. *Phys. Rev. B* **91**, 115405 (2015).

28. Yin, L.-J., Qiao, J.-B., Yan, W., Xu, R., Dou, R.-F., Nie, J.-C., He, L. Electronic structures and their Landau quantizations in twisted graphene bilayer and trilayer. arXiv:1410.1621.

**Acknowledgements**

This work was supported by the National Basic Research Program of China (Grants Nos. 2014CB920903, 2013CBA01603), the National Natural Science Foundation of China (Grant Nos. 11422430, 11374035), the program for New Century Excellent Talents in University of the Ministry of Education of China (Grant No. NCET-13-0054), Beijing Higher Education Young Elite Teacher Project (Grant No. YETP0238).

**Author contributions**

L.H. conceived and provided advice on the experiment, analysis, and theoretical calculation. L.J.Y. performed the STM experiments and analyzed the data. J.B.Q., W.J.Z., and W.T.L. performed the theoretical calculations. L.H. wrote the paper. All authors participated in the data discussion.

**Competing financial interests:** The authors declare no competing financial interests.

**Figure Legends**

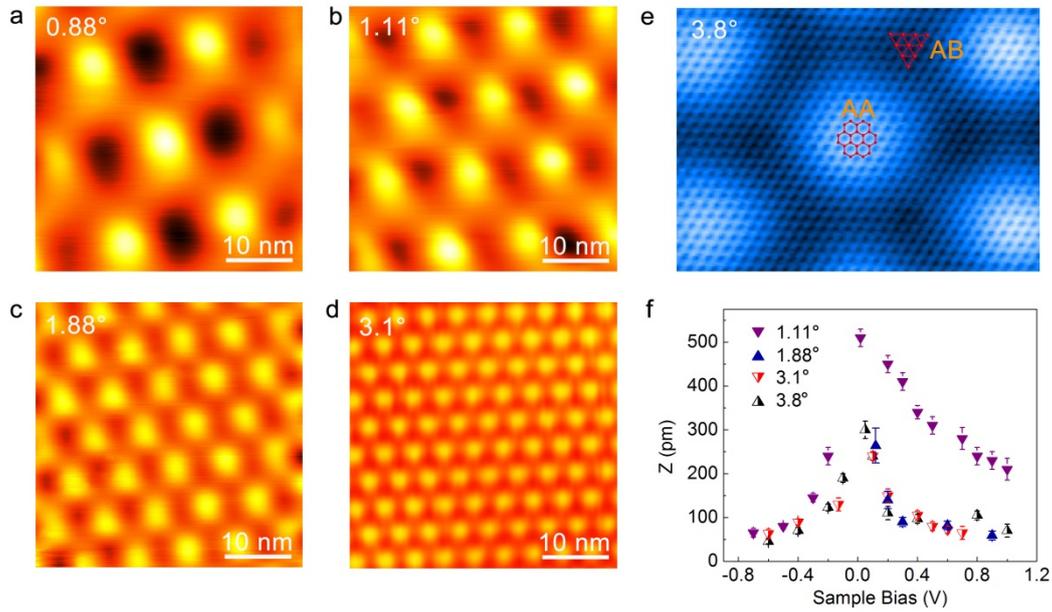

**Figure 1 | STM topographs of twisted graphene bilayers. a-d,** STM topographic images, 40 nm × 40 nm, of twisted bilayers with the twisted angle $\theta = (0.88\pm0.03)°$ and the period of the Moiré patterns $D = (16.0\pm0.5)$ nm (**a**), $\theta = (1.11\pm0.05)°$ and $D = (12.6\pm0.6)$ nm (**b**), $\theta = (1.88\pm0.08)°$ and $D = (7.5\pm0.3)$ nm (**c**), and $\theta = (3.1\pm0.1)°$ and $D = (4.5\pm0.3)$ nm (**d**) on graphite surface. **e,** Atomic resolution STM image of a twisted bilayer with $\theta = (3.8\pm0.1)°$ and $D = (3.7\pm0.1)$ nm. It shows different stacking orders depending on positions of the moiré structure. In the AA-stacking region, we can observe honeycomb lattice, whereas, in the AB-stacking region, only triangular lattice can be observed. **f,** Range of STM measured corrugation of different moiré patterns as a function of sample voltages.

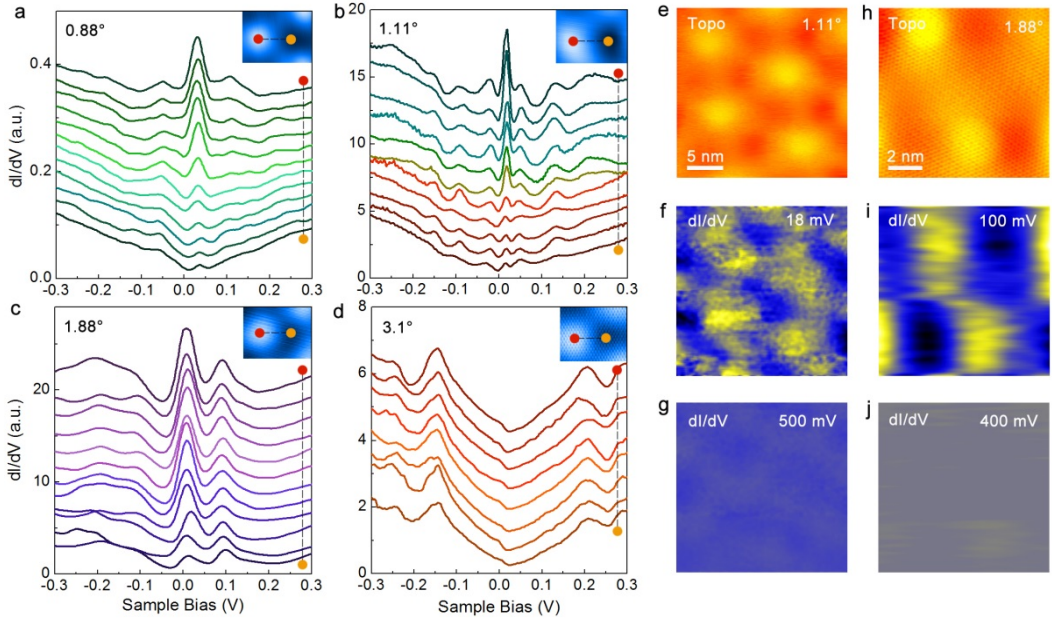

**Figure 2 | STS spectra and d*I*/d*V* maps of twisted graphene bilayers. a-d,** Spatial evolution of d*I*/d*V* spectra along the dashed lines in STM images of the insets for the twisted bilayers with $\theta = (0.88\pm0.03)°$ (**a**), $\theta = (1.11\pm0.05)°$ (**b**), $\theta = (1.88\pm0.08)°$ (**c**), and $\theta = (3.1\pm0.1)°$ (**d**). For the 1.88° and 3.1° bilayers, the two peaks with energy spacing of 80 meV and 350 meV, respectively, are the two low-energy VHSs. For the 1.11° and 0.88° samples, a pronounced peak mainly localized in the AA-stacking regions is observed at the charge neutrality point of the spectra, indicating the emergence of nondispersive flat bands. The FWHMs of the peaks for the 1.11° and 0.88° bilayers are 18 meV and 30 meV, respectively. **e,** 25 nm × 25 nm STM topography image of the 1.11° moiré patterns. **f** and **g**, d*I*/d*V* maps of the same region in **e** at two bias voltages. The map recorded at 18 mV, the energy of the flat bands, shows localized states following the modulation of the Moiré patterns, whereas, the map recorded at 500 mV, the energy far from the flat bands, exhibits extended states. **h,** 10 nm × 10 nm STM topography image of the 1.88° moiré patterns. **i** and **j**, d*I*/d*V*

maps of the same region in **h** at two bias voltages, 100 mV (**i**) and 400 mV (**j**). It also displays localization of the DOS in the d*I*/d*V* map recorded at 100 mV, the position of one of the VHSs in panel **c**. The charge density becomes nearly homogeneous for energies away from the VHSs.

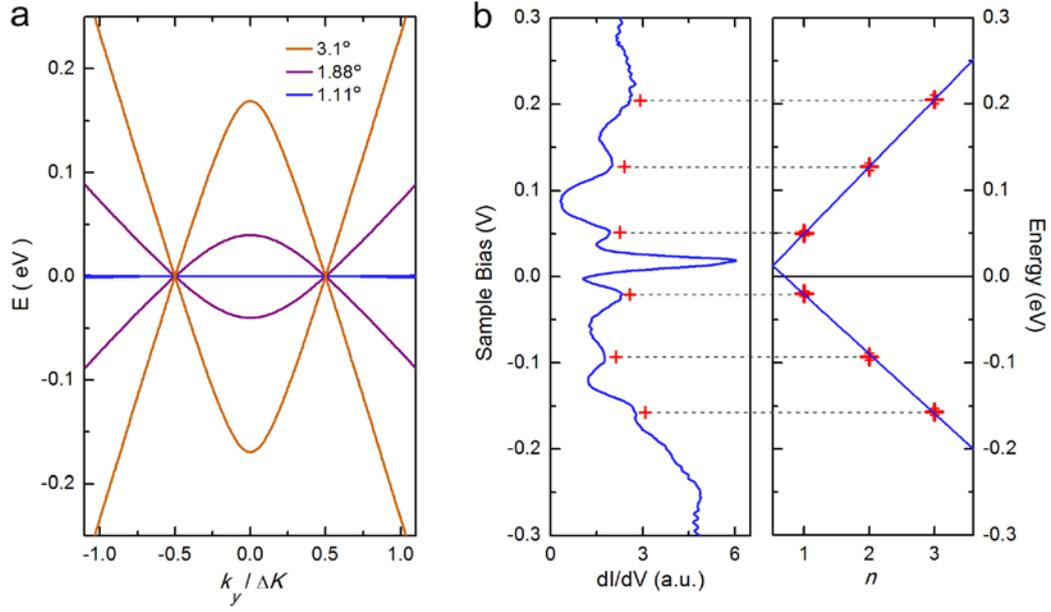

**Figure 3 | Electronic band structures and quantum confinement in twisted bilayers. a,** Low-energy sub-band of twisted bilayers with different twisted angles. The energy separation of the two saddle points decreases with decreasing the rotation angle and the low-energy nondispersive flat bands appear in the 1.11° bilayer. For clarity, only the lowest subbands are shown. **b,** Left panel: A typical differential conductance spectrum recorded in the 1.11° bilayer. Besides the pronounced peak at the charge neutrality point, several regularly spaced resonances appear in the spectrum, as marked by the red crosses. Right panel: Energy positions of the resonance peaks in the left panel as a function of the positive integers. The energy separation of these peaks is about 70 meV.

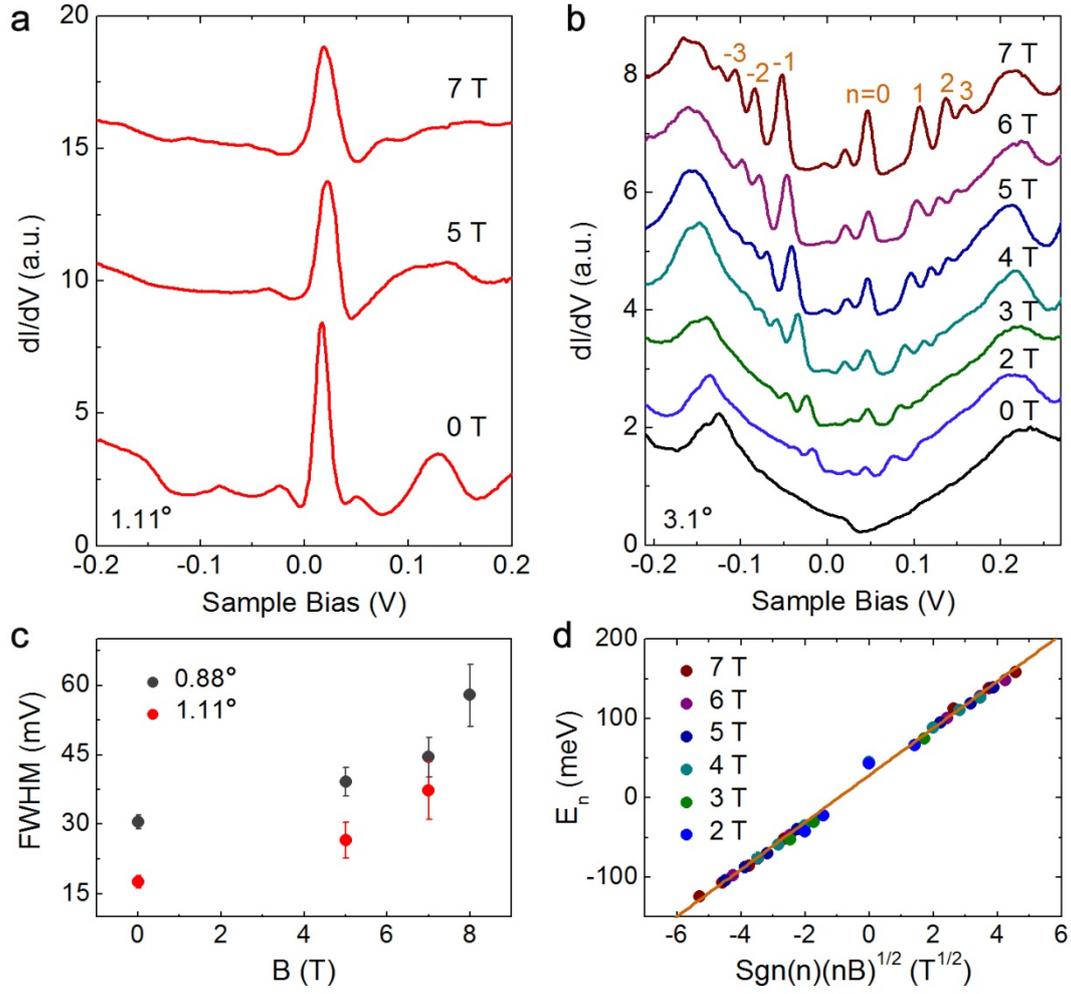

**Figure 4 | STS spectra of twisted graphene bilayers in different magnetic fields. a,** d$I$/d$V$ spectra taken at a bright dot in the 1.11° bilayer with different magnetic fields. **b,** Tunnelling spectra of the 3.1° moiré pattern measured under various magnetic fields. For clarity, the curves are offset in Y-axis and the Landau level indices of graphene monolayer are marked. **c,** FWHM of the peaks at the charge neutrality points of the 0.88° and 1.11° moiré patterns in different magnetic fields. **d,** Landau level peak energies for different magnetic fields obtained in panel **b** are shown to be linear with sgn($n$)($|n|B$)$^{1/2}$, as expected for massless Dirac fermions. The solid line is a linear fitting of the data with the slope yielding a Fermi velocity of $v_F = (0.82 \pm 0.01) \times 10^6$ m/s.